\documentstyle[12pt,a4]{article} 
\begin{document}
\thispagestyle{empty}
\renewcommand{\thefootnote}{\fnsymbol{footnote}}
\begin{flushright}
{\tt DPNU-99-04 \\
February 1999}
\end{flushright}
\baselineskip 5mm

\begin{center}
{\LARGE Exact Parametrization of Majorana Neutrino \\
\vspace*{2mm}
Mass Matrix with Large Mixing} \\
\vspace*{5mm}
{\large {\sc K. Kimura}
 {\sc $\;$and  A. Takamura\footnote{
                e-mail address:
                kimukei@eken.phys.nagoya-u.ac.jp; 
                takamura@eken.phys.nagoya-u.ac.jp
            }} \\
\vspace*{2mm}
 Department of Physics, Nagoya University \\
\vspace*{1mm}
     Nagoya 464-8602, Japan}
\end{center}
\date{}

\bigskip

\begin{abstract}
Under the assumption that the neutrinos are Majorana particles 
we study how the lepton mass matrices can be transformed into the simple 
form which has the same physical quantities 
by removing redundant parameters.
We propose the exact parametrization of 
the lepton mass matrices which reflects the small $\nu_e-\nu_{\mu}$ mixing 
and the large $\nu_{\mu}-\nu_{\tau}$ mixing.
The relations between the twelve parameters 
and the physical quantities are shown.
Furthermore we calculate the MNS matrix by applying the assumptions
used in the quark sector.  
Finally we also check the validity of these assumptions 
from the experimental values.
\end{abstract}

\section{Introduction}

\hspace*{\parindent}
The finite mass of neutrinos and the mixing among different 
flavor neutrinos have been suggested by various neutrino 
oscillation experiments.
One of the major experiments is the solar neutrino experiments, 
which indicate the oscillation between $\nu_e$ and other neutrinos.
Another one is the atmospheric neutrino experiments, which 
indicate the oscillations between $\nu_{\mu}$ and other neutrinos. 

In solar neutrino experiments three possible solutions are proposed:
small or large mixing MSW solution \cite{MSW} and vacuum oscillation solution 
with large mixing \cite{VO}.
Especially the small mixing MSW solution \cite{SMSW}
\begin{equation}
\sin^2 {2 \theta_{ex}} \sim 10^{-2},\, 
\Delta m^2_{\rm solar} \sim 10^{-5} \,{\rm eV}^2 \label{43},
\end{equation}
has been thought as the strong candidate compared with others 
for the solar neutrino problem.
Furthermore $\nu_e \rightarrow \nu_{\mu}$ oscillations 
are the most likely channel in (\ref{43}) 
although other channels
$\nu_e \rightarrow \nu_{\tau}$ and $\nu_e \rightarrow \nu_s$ 
are not be excluded.

Another experiments are the atmospheric neutrino experiments.
In particular, the recent report by Super-Kamiokande \cite{K.F}
concerned with a zenith-angle-dependent deficit of $\nu_{\mu}$ 
suggests the strong evidence for neutrino oscillation 
with large mixing,
\begin{equation}
\sin^2 {2 \theta_{\mu x}} \sim 1,\,
\Delta m_{\rm atm}^2 \sim 10^{-3} \, {\rm eV}^2.
\end{equation}
Within the three neutrino picture, 
the atmospheric neutrino problem can be explained by 
$\nu_{\mu} \rightarrow \nu_{\tau}$ because 
$\nu_{\mu} \rightarrow \nu_e$ is excluded 
by the CHOOZ experiment \cite{CHOOZ}. 
In the near future it is expected that the data obtained 
by Super-Kamiokande, SNO, K2K and so on reveal 
the structure of the lepton sector more clearly.
 
At the moment, we recognize that the lepton sector is 
largely different from the quark sector in following two points.
The first one is the large $\nu_{\mu}-\nu_{\tau}$ mixing 
in contrast to quark sector.
The second one is that the neutrino masses are extremely small compared
with the quark and the charged lepton masses. 

These differences seem to be explained by more fundamental theory.
Actually, 
the smallness of neutrino masses can be naturally understood 
as the inverse of large Majorana neutrino masses  
by using seesaw mechanism \cite{Y}.
We can also consider that the large mixing 
is not originated from the charged lepton but from the Majorana neutrino.
Because it is expected that 
the charged lepton mass matrix has the same structure 
as the quark one in the grand unified theory (GUT). 
In addition, it is expected that 
the contribution of the charged lepton 
to mixing angle is small at weak scale as well as GUT scale 
since the renormalization effects are small in general.
 
As one of the mass matrices which leads the small mixing, 
Fritzsch-type \cite{F} mass matrix is proposed in the quark sector.
It is given by imposing hermicity 
to the nearest-neighbor interactions (NNI) form, 
which has the components $M_{11}=M_{13}=M_{31}=M_{22}=0$.
One of approaches to explore the symmetry of more fundamental theory 
is to consider the simple mass matrices like this.
In general Branco, Lavoura and Mota \cite{B.L.M} 
proved that one can always choose the NNI form 
as both up and down quark mass matrices.
This means that arbitrary mass matrices can be transformed into the 
NNI form which has the same physical quantities.

In this letter inspired by these works, 
we explore the simple form of Majorana neutrino mass matrix, $M_{\nu}$, 
which only contributes to the $\nu_{\mu}-\nu_{\tau}$ mixing 
when we choose the NNI form as the charged lepton mass matrix $M_l$.
As a result, 
we prove that one can always choose the following simple forms
as Majorana neutrino mass matrix, 
\begin{equation}
M_{\nu} \propto
\left(
 \begin{array}{ccc}
     a_{\nu} & 0 & 0 \\
     0 & b_{\nu} & d_{\nu} \\
     0 & d_{\nu} & c_{\nu} 
 \end{array}
\right), \label{1}
\end{equation} 
where $a_{\nu}, b_{\nu}, c_{\nu}$ and $d_{\nu}$ are 
complex values and are represented by five independent parameters,
and charged lepton mass matrix, 
\begin{equation}
M_l \propto
\left(
 \begin{array}{ccc}
     1 & 0 & 0 \\
     0 & \exp(i \theta_2) & 0 \\
     0 & 0 & \exp(i \theta_3) 
 \end{array}
\right)
\times 
\left(
 \begin{array}{ccc}
     0 & a & 0 \\
     c & 0 & b \\
     0 & d & e 
 \end{array}
\right) \label{2},
\end{equation}
where $a,b,c,d$ and $e$ are all real values. 
Twelve parameters are contained in $M_l$ and $M_{\nu}$ and 
just correspond to the physical quantities: 
three charged lepton masses, three neutrino masses, 
three mixing angles and three phases.
Note that arbitrary $M_{\nu}$ and $M_l$ can be transformed 
into the simple forms (\ref{1}) and (\ref{2}) 
which have the same physical quantities. 

These (\ref{1}) and (\ref{2}) are the exact parametrization 
of the lepton mass matrices.
In this parametrization, there is no contributions to 
$\nu_e-\nu_{\mu}$ mixing from $M_{\nu}$ 
but only from $M_l$.  
On the other hand, $\nu_{\mu}-\nu_{\tau}$ mixing is contributed from 
both $M_l$ and $M_{\nu}$. 
The large mixing can be generated only by $M_{\nu}$ 
because the contribution from $M_l$ is small.
Thus, this parametrization is suitable for the physics
with the small mixing MSW solution for solar neutrino problem 
and the large mixing solution for atmospheric neutrino problem.

In the following section we give the proof how to 
transform into (\ref{1}) and  (\ref{2}).
Next we start from (\ref{1}) and (\ref{2}),
and calculate the Maki-Nakagawa-Sakata (MNS) matrix 
\cite{M.N.S}, which represents lepton mixing matrix,  
under the assumptions used in the quark sector. 
Finally we check the validity of these assumptions from the 
experimental values.

\section{Simple Form of Lepton Mass Matrices}

\hspace*{\parindent}
In this section, 
starting from the arbitrary Majorana neutrino mass matrix $M_{\nu}$ 
and the charged lepton Dirac mass matrix $M_l$,
we transform the neutrino mass matrix into the simple form 
(\ref{1}) and the charged lepton mass matrix 
into the NNI form (\ref{2}) by using the following transformation
\begin{eqnarray}
M_{\nu}^{\prime}&=&U^T M_{\nu} U, \label{3}\\
M_l^{\prime}&=&U^{\dagger} M_l V. \label{4}
\end{eqnarray}
The physical quantities calculated by $(M_{\nu},M_l)$ are not changed
through the above transformation to $(M_{\nu}^{\prime},M_l^{\prime})$.
In general $M_{\nu}$ is a $3 \times 3$ complex symmetric matrix, 
which has 12 parameters and $M_l$ is a $3 \times 3$ complex matrix, 
which has 18 parameters.
Taking advantage of freedom contained in $U$ and $V$, 
the number of the parameters in $M_l$ and $M_{\nu}$ is reduced 
from 30 to 12. 
The remaining 12 parameters in $M_l$ and $M_{\nu}$ just 
correspond to the physical quantities.

At first let us show how to transform $M_l$ into NNI form, 
which has the components 
$(M_l^{\prime})_{11}=(M_l^{\prime})_{13}=(M_l^{\prime})_{31}
= (M_l^{\prime})_{22}=0$,
based on the work done by Branco {\it et al}. \cite{B.L.M}.
Three conditions 
$(M_l^{\prime})_{11}=(M_l^{\prime})_{13}=(M_l^{\prime})_{31}=0$ 
are satisfied by choosing the unitary matrix 
$V=(V_{i1},V_{i2},V_{i3})$ in (\ref{4}) as
\begin{eqnarray}
V_{i1} &=& N_1 \epsilon_{ijk}{\cal M}_{1j}{\cal M}_{3k}, \label{27}\\
V_{i3} &=& N_3 ({\cal M}_{1i}^*(H_l^{\prime})_{13}
-{\cal M}_{3i}^*(H_l^{\prime})_{11}) \label{28},
\end{eqnarray}
where the $N_1$ and $N_3$ are normalization factors
and ${\cal M}$, $H_l^{\prime}$ are defined by 
\begin{eqnarray}
{\cal M} &\equiv& U^{\dagger}M_l, \label{32}\\
H_l^{\prime} &\equiv& U^{\dagger} H_l U=
U^{\dagger} M_l M_l^{\dagger} U={\cal M} {\cal M}^{\dagger} \label{29}.
\end{eqnarray} 
Note that $U$ is an arbitrary unitary matrix in (\ref{32}) and 
(\ref{29}).
The vector $V_{i2}$ is  
determined by the orthogonality conditions 
with the vectors $V_{i1}$ and $V_{i3}$ as 
\begin{equation}
V_{i2} = N_2 {\cal M}_{1i}^* \label{30},
\end{equation}
where the $N_2$ is an also normalization factor. 
After the transformations (\ref{27}) and (\ref{28}), 
the matrix element $(M_l^{\prime})_{22}$ turns to  
\begin{equation}
 (M_l^{\prime})_{22}=({\cal M})_{2i} V_{i2}=
N_2 (H_l^{\prime})_{21},
\end{equation}
in terms of (\ref{29}) and (\ref{30}).
The remaining condition for the NNI form is realized 
in $(H_l^{\prime})_{21}=0$.
For the purpose, we impose $U$ to satisfy the following condition, 
\begin{equation}
(H_l^{\prime})_{21}=U_{i2}^*(H_l)_{ij}U_{j1}=0 
\label{26}.
\end{equation}

Next we transform the neutrino mass matrix $M_{\nu}$  
into the simple form (\ref{1}). 
This is done by choosing $U$ in (\ref{3}) as   
\begin{eqnarray}
& &(M_{\nu}^{\prime})_{21}=(M_{\nu}^{\prime})_{12}=
U_{i1}(M_{\nu})_{ij}U_{j2}=0, \label{24} \\
& &(M_{\nu}^{\prime})_{31}=(M_{\nu}^{\prime})_{13}=
U_{i1}(M_{\nu})_{ij}U_{j3}=0 \label{25}.
\end{eqnarray}
These (\ref{24}) and (\ref{25}) are contended 
if $U$ is satisfied in the following equation   
\begin{equation}
U_{i1}^*(M_{\nu})_{ij}^*=N_4 U_{j1} \label{8},
\end{equation}
where $N_4$ is an normalization factor. 
The solution $U_{i1}$ in this equation 
is obtained by $U=(U_{i1},U_{i2},U_{i3})$ 
which diagonalize $M_{\nu}^{\dagger}M_{\nu}$. 

Once $U_{i1}$ is obtained, $U_{i2}$ is successively determined 
by the condition (\ref{26}), 
and $U_{i3}$ is also determined by the orthogonality conditions 
with the other two vectors $U_{i1}$ and $U_{i2}$.
The explicit forms of $U_{i2}$ and $U_{i3}$ are written with $U_{i1}$ as 
\begin{eqnarray}
 U_{i2}&=& N_5 \epsilon_{ijk} U_{j1}^* 
(H_l^* U^*)_{k1}, \label{7} \\
U_{i3}&=&N_6 ((H_l)_{ij}U_{j1}-(H_l^{\prime})_{11} U_{i1}) \label{6},
\end{eqnarray}
where $N_5$ and $N_6$ are normalization factors.

Finally we show that the number of parameters are reduced to 
12 by the ambiguity of the six phases containing in the normalization 
factors of eigenstates $U_{i1},U_{i2},U_{i3}$ and 
$V_{i1},V_{i2},V_{i3}$.
The redefinitions $U \to U P^{\dagger}(\delta_{\nu})$ 
and $V \to V P^{\dagger}(\delta_{R})$ 
do not change the forms of lepton mass matrices (\ref{1}) and (\ref{2}), 
where $P(\delta_{\nu})$ and $P(\delta_R)$ are diagonal phase matrices.
By these redefinitions, (\ref{3}) and (\ref{4}) become 
\begin{eqnarray}
M_{\nu}^{\prime\prime}&=&
P^{\dagger}(\delta_{\nu})U^TM_{\nu}UP^{\dagger}(\delta_{\nu})
=P^{\dagger}(\delta_{\nu})M_{\nu}^{\prime}P^{\dagger}(\delta_{\nu}), \\
M_l^{\prime\prime}&=&
P(\delta_{\nu})U^{\dagger}M_lVP^{\dagger}(\delta_R)
=P(\delta_{\nu})M_l^{\prime}P^{\dagger}(\delta_R) \label{31}.
\end{eqnarray}
We can choose $P(\delta_{\nu})$ so as to  
absorb three phases of $M_{\nu}^{\prime}$   
and as a result one phase degree of freedom is only 
left in $M_{\nu}^{\prime\prime}$.
Introducing another phase matrix $P(\delta_L)$ 
to separate the phase factor from $M_l^{\prime}$, 
we can rewrite (\ref{31}) as 
\begin{eqnarray}
M_l^{\prime\prime} &=& P(\delta_{\nu})P^{\dagger}(\delta_L)\times
P(\delta_L)U^{\dagger}M_lVP^{\dagger}(\delta_R) \\
&=&P(\delta_{\nu})P^{\dagger}(\delta_L)\times
P(\delta_L)M_l^{\prime}P^{\dagger}(\delta_R) \\
&=& P(\delta)\hat{M_l},
\end{eqnarray}
where $P(\delta) \equiv P(\delta_{\nu})P^{\dagger}(\delta_L)$ is 
a diagonal phase matrix and 
${\hat M} \equiv P(\delta_L)U^{\dagger}M_lVP^{\dagger}(\delta_R)$ 
is a real symmetric matrix of NNI form \cite{F,I.T}.
$P(\delta_L)$ and $P(\delta_R)$ can be chosen so that 
five phases of $M_l^{\prime}$ are absorbed.

By replacement $M_{\nu}^{\prime\prime}$ with $M_{\nu}$ 
and $M_l^{\prime\prime}$ with $M_l$, 
we can always choose 
the lepton mass matrices as  
\begin{eqnarray}
M_{\nu}&=& m_{\nu_3}
\left(
 \begin{array}{ccc}
     a_{\nu} & 0 & 0 \\
     0 & b_{\nu} & d_{\nu} \\
     0 & d_{\nu} & c_{\nu} 
 \end{array}
\right) \label{12}, \\
M_l&=&P(\delta)\hat{M_l} \label{35} \\
&=& 
\left(
 \begin{array}{ccc}
     1 & 0 & 0 \\
     0 & \exp(i \theta_2) & 0 \\
     0 & 0 & \exp(i \theta_3) 
 \end{array}
\right)
\times m_3
\left(
 \begin{array}{ccc}
     0 & a & 0 \\
     c & 0 & b \\
     0 & d & e 
 \end{array}
\right) \label{11},
\end{eqnarray}
where $m_{\nu_3}$ and $m_3$ are respectively the heaviest neutrino mass 
and the heaviest charged lepton mass and $M_{\nu}$ 
is complex symmetric including only one phase 
degree of freedom implicitly in some components.
In (\ref{11}) we make $(P(\delta))_{11}=1$ using 
the remaining freedom in $P(\delta_L)$.
Note that $M_{\nu}$ and $M_l$ can be chosen as (\ref{12}) and (\ref{11}) 
at any scale.
As mentioned in the introduction, the above representation 
(\ref{12}) of $M_{\nu}$  only contributes to 
the $\nu_{\mu}-\nu_{\tau}$ mixing. 
The lepton mass matrices (\ref{12}) and (\ref{11}) have 
twelve parameters including three phases and these parameters 
are just the same as the number of physical quantities.

\section{Physical Quantities and Twelve Parameters}

\hspace*{\parindent}
In this section, 
our purpose is to study the relations between 
the physical quantities and twelve parameters 
for the charged lepton mass matrix $M_l$ of (\ref{12}) and 
the neutrino mass matrix $M_{\nu}$ of (\ref{11}). 
The mass matrices (\ref{12}) 
and (\ref{11}) should generate correct mass eigenvalues.
Introducing the three charged lepton mass eigenvalues 
for $\hat{M_l}$ in (\ref{35}) as input parameters, 
$M_l$, which has five free parameters at first, is parametrized by 
two free parameters.
In the same way, $M_{\nu}$ is also parametrized by two free 
parameters. 
Then, 
the remaining six free parameters (each of ${\hat M_l,M_{\nu}}$ and 
$P(\delta)$ has two parameters) can be determined by the 
MNS matrix denoted by $V_{MNS}$,
which should be fixed by the physical quantities, 
\begin{equation}
V^{\dagger}_{MNS}=U_{\nu}^{\dagger}P(\delta)O_l,
\end{equation}
where $U_{\nu}$ is the unitary matrix which diagonalize $M_{\nu}$ 
and $O_l$ is the orthogonal matrix which diagonalize $\hat{M_l}\hat{M_l}^T$.

At first, let us consider the charged lepton mass matrix $M_l$.
We apply the work for the quark sector done 
by Harayama and Okamura \cite{H.O} to the lepton sector.
$\hat{M_l} \hat{M_l}^T$ is a real symmetric matrix 
which can be diagonalized by orthogonal 
matrix $O_l$ as 
$O_l^T \hat{M_l} \hat{M_l}^T O_l= 
m_3^2{\rm diag\/}(\xi_1^2,\xi_2^2,\xi_3^2)$,
where $\xi_1 \equiv m_1/m_3,\, \xi_2 \equiv m_2/m_3 \,
\xi_3 \equiv m_3/m_3=1$ and $m_i(i=1,2,3)$ are the charged lepton masses.

As the results, we can parametrize 
the charged lepton mass matrix as 
\begin{equation}
{\hat M_l}=m_3
\left(
 \begin{array}{ccc}
     0 & qz/y & 0 \\
     q/(yz) & 0 & b \\
     0 & d & y^2 
 \end{array}
\right) \label{15},
\end{equation}
where the matrix elements $b$ and $d$ are expressed as 
\begin{eqnarray}
b&=&\sqrt{\frac{p+1-y^4 \pm R}{2} - \frac{q^2}{y^2 z^2}} \label{19},\\
d&=&\sqrt{\frac{p+1-y^4 \mp R}{2} - \frac{q^2 z^2}{y^2}} \label{20},
\end{eqnarray}
and 
\begin{equation}
R = \sqrt{(1+p-y^4)^2-4(p+q^4)+4q^2 y^2 
\left( z^2 + \frac{1}{z^2} \right)},
\end{equation}
\begin{eqnarray}
q^2&=&\xi_1 \xi_2, \\
p&=&\xi_1^2+\xi_2^2.
\end{eqnarray}
Thus, $M_l$ has only two independent free parameters 
$y$ and $z$.

The orthogonal matrix, $O_l$, which diagonalize 
$\hat{M_l} \hat{M_l}^T$ 
is also represented using only two independent free parameters 
as follows; 
\begin{equation}
 O_l=
 \left(
 \begin{array}{c}
     \alpha_i \\
     \beta_i \\
     \gamma_i 
 \end{array}
\right) \\
=\frac{1}{f_i}
 \left(
 \begin{array}{c}
     (\xi_i^2-b^2-c^2)ad \\
     (\xi_i^2-a^2)be   \\
     (\xi_i^2-a^2)(\xi_i^2-b^2-c^2) 
 \end{array}
 \right) \label{13},
\end{equation}
where $a=qz/y, c=q/(yz)$ and $f_i$ is normalization factor.

Next we consider the neutrino mass matrix $M_{\nu}$.
$M_{\nu}^{\dagger}M_{\nu}$ 
should generate correct mass eigenvalues as 
$U_{\nu}^{\dagger} M_{\nu}^{\dagger}M_{\nu} U_{\nu}=D_{\nu}^2=
m_{\nu_3}^2{\rm diag\/}
(\xi_{\nu_1}^2,\xi_{\nu_2}^2,\xi_{\nu_3}^2)$ by the 
unitary matrix $U_{\nu}$, 
where $\xi_{\nu_1} \equiv m_{\nu_1}/m_{\nu_3},\,
\xi_{\nu_2} \equiv m_{\nu_2}/m_{\nu_3},\,
\xi_{\nu_3} \equiv m_{\nu_3}/m_{\nu_3} =1$ and 
$m_{\nu_i}(i=1,2,3)$ are the neutrino masses.
In general the matrix $M_{\nu}^{\dagger}M_{\nu}$ can be 
parametrized without loss of generality as follows;
\begin{equation}
  M_{\nu}^{\dagger}M_{\nu}=m_{\nu_3}^2
\left(
 \begin{array}{ccc}
     \xi_{\nu_1}^2 & 0 & 0 \\
     0 & 1-C+\xi_{\nu_2}^2 & 
-\sqrt{(1-C)(C-\xi_{\nu_2}^2)} \, e^{-i\theta_1}
\\
     0 & -\sqrt{(1-C)(C-\xi_{\nu_2}^2)} \, e^{i\theta_1} & C 
 \end{array}
\right)\label{21},
\end{equation}
using two parameters $C$ and $\theta_1$ 
so that trace and determinant are invariant under unitary transformations.
The above matrix can be diagonalized by 
\begin{equation}
 U_{\nu}=
\left(
 \begin{array}{ccc}
     1 & 0 & 0 \\
     0 & e^{-i\theta_1}\cos \phi & -\sin \phi   \\
     0 & \sin \phi & e^{i\theta_1}\cos \phi  
 \end{array}
\right) \label{33},
\end{equation}
where 
\begin{equation}
 \cos \phi=\sqrt{\frac{C-\xi_{\nu_2}^2}{1-\xi_{\nu_2}^2}},\,
 \sin \phi=\sqrt{\frac{1-C}{1-\xi_{\nu_2}^2}}.
\end{equation}

The neutrino mass matrix $M_{\nu}$ reconstructed from $U_{\nu}$ 
and $D_{\nu}$ is 
\begin{eqnarray}
M_{\nu}&=& U_{\nu}^* D_{\nu} U_{\nu}^{\dagger} \nonumber \\
&=& m_{\nu_3}
\left(
 \begin{array}{ccc}
     \xi_{\nu_1} & 0 & 0 \\
     0 & A-B\cos 2\phi & -e^{-i\theta_1}B\sin 2\phi  \\
     0 & -e^{-i\theta_1}B\sin 2\phi 
& e^{-2i\theta_1}(A+B\cos 2\phi)
 \end{array}
\right) \label{16},
\end{eqnarray}
where 
\begin{equation}
 A=\frac{1}{2}(1+\xi_{\nu_2}e^{2i\theta_1}),
 B=\frac{1}{2}(1-\xi_{\nu_2}e^{2i\theta_1}).
\end{equation}
(\ref{15}) and (\ref{16}) are the exact parametrization of 
the lepton mass matrices.
These simple form can be obtained without any approximations.

Finally we obtain the MNS matrix by using 
(\ref{11}), (\ref{13}) and (\ref{33}) as follows;
\begin{equation}
V^{\dagger}_{MNS}=U_{\nu}^{\dagger}P(\delta)O_l =
\left(
 \begin{array}{c}
  \alpha_i \\
  \beta_i \cos \phi \, e^{i(\theta_1+ \theta_2)} + 
\gamma_i \sin \phi \, e^{i\theta_3} \\
  -\beta_i \sin \phi \, e^{i \theta_2} + 
\gamma_i \cos \phi \, e^{i (\theta_3-\theta_1)} \\
 \end{array}
 \right) \label{17},
\end{equation}
where $\alpha_{i}$, $\beta_{i}$ and $\gamma_{i}$ are expressed 
by $y$ and $z$ in (\ref{13}).

In (\ref{17}), six independent parameters,
$y,z,\phi,\theta_1,\theta_2, \theta_3$, are included.
$y$, $z$, $\phi$ correspond to three mixing angles 
and $\theta_1,\theta_2, \theta_3$ correspond to three CP phases.
If the contribution of the charged lepton mass matrix to 
$\nu_{\mu}-\nu_{\tau}$ mixing is small, we can identify $\phi$ 
with the large $\nu_{\mu}-\nu_{\tau}$ mixing angle.

\section{The MNS Matrix}

\hspace*{\parindent}
In this section, 
we investigate the structure of the MNS matrix
under the approximation that the mass ratios of the charged leptons 
$\xi_1 (= m_1/m_3)$ and $\xi_2 (= m_2/m_3)$ are small quantities 
compared with $\xi_3= 1 (= m_3/m_3)$.
In addition, we adopt the assumptions $y \sim O(1), z \sim O(1)$ 
in the charged lepton mass matrix $M_l$, 
based on the expectation that it has the same structure 
as the quark mass matrices, as mentioned in the introduction.
We roughly estimate the values of $y$ and $z$ 
from the experimental values and check the validity of 
the above assumptions. 
Then, we also study the  structure of neutrino mass matrix 
$M_{\nu}$ under the condition that the maximal $\nu_{\mu}-\nu_{\tau}$ 
mixing is derived from $V_{MNS}$,
in each case that neutrinos have hierarchical masses 
($\xi_{\nu_1} \ll \xi_{\nu_2} \ll 1$) or degenerate masses
($\xi_{\nu_1} \sim \xi_{\nu_2} \sim 1$).

Let us start from the exact form of (\ref{15}). 
The mass hierarchy of the charged leptons, 
$\xi_1 \ll \xi_2 \ll 1$, 
and the assumptions, $y \sim O(1)$ and $z \sim O(1)$, 
lead to the results,  
\begin{equation}
 R \sim (1-y^4)-\frac{1+y^4}{1-y^4}p \label{18},
\end{equation}
and then substituting (\ref{18}) for $R$ in (\ref{19}) and (\ref{20}), 
we obtain either case I or case II, 
which is corresponding to the sign in front of $R$ 
in (\ref{19}) and (\ref{20}), 
\begin{eqnarray}
 {\rm case~I}&:& b \sim \sqrt{1-y^4} \sim O(1), 
 \:\: d \sim \sqrt{\frac{p}{1-y^4}} \ll O(1), \\
 {\rm case~II}&:& b \sim \sqrt{\frac{p}{1-y^4}} \ll O(1),
 \:\: d \sim \sqrt{1-y^4} \sim O(1) \label{39}.
\end{eqnarray}

The case I and II have quite different structures at the following point.
In the case I, $d$ is a small quantity compared with $b$,
oppositely in the case II, $b$ is a small quantity compared with $d$. 
This leads to a different consequence for the mixing angle. 
Actually it is shown by Ref. \cite{H.O.S.X} 
that the mass matrix leads to large mixing 
between the second and the third generations in the case I, 
in contrast it leads to small mixing in the case II.
The case I is not consistent with our assumption that 
the charged lepton mass matrix has only small mixing. 
Therefore in the following discussion we only treat the case II and 
do not describe a detail calculation for the case I.

Next we adopt the assumption (\ref{39}) in the MNS matrix 
$V^{\dagger}_{\rm MNS}$ of (\ref{17}) and we obtain  
\begin{equation}
V^{\dagger}_{\rm MNS}\simeq
\left(
 \begin{array}{ccc}
     1 & qyz/\sqrt{p} & -qz\sqrt{1-y^4}/y \\
     - qyz\cos \phi \, e^{i(\theta_1+\theta_2)}/\sqrt{p} & 
\cos \phi \, e^{i(\theta_1+ \theta_2)} & 
\sin \phi \, e^{i \theta_3} \\
     qyz\sin \phi \, e^{i \theta_2}/\sqrt{p} & 
-\sin \phi \, e^{i \theta_2} & 
\cos \phi \, e^{i (\theta_3-\theta_1)}
\end{array}
\right)\label{22},
\end{equation}
at leading order approximation.
Here we roughly estimate the values of $y$ and $z$ from the 
experimental values and check the validity of the assumptions 
$y \sim O(1)$ and $z \sim O(1)$.
Introducing the experimental values, 
\begin{eqnarray}
m_e(M_Z) &\sim& 0.4867 \,{\rm MeV}, \\
m_{\mu}(M_Z) &\sim& 102.7 \,{\rm MeV}, \\ 
m_{\tau}(M_Z) &\sim& 1747 \,{\rm MeV},
\end{eqnarray}
and $\phi=45^{\circ}$ for the maximal $\nu_{\mu}-\nu_{\tau}$ mixing, 
(\ref{22}) becomes the following;
\begin{equation}
V^{\dagger}_{\rm MNS}\simeq
\left(
 \begin{array}{ccc}
     1 & 0.0689yz & -0.00405z\sqrt{1-y^4}/y \\
     -0.0487yz & 0.707 & 0.707 \\
     0.0487yz & -0.707 & 0.707
\end{array}
\right) \label{40},
\end{equation}
up to phase factor of each matrix element.
If $y$ and $z$ are fixed, we can determine the MNS matrix.
As one of the examples, 
we obtain the MNS matrix in the case $y=0.90$ and $z=0.62$;
\begin{equation}
V^{\dagger}_{\rm MNS}\simeq
\left(
 \begin{array}{ccc}
     1 & 0.038 & -0.002 \\
     -0.027 & 0.707 & 0.707 \\
     0.027 & -0.707 & 0.707
\end{array}
\right).
\end{equation}
This result is consistent with the 
best fit values $\theta_{e\mu} \sim 2.2^{\circ}$ and 
$\theta_{e\tau} \sim 0^{\circ}$ \cite{CHOOZ,G.N.P.V}. 
Therefore we recognize that the assumptions $y \sim O(1)$ and 
$z \sim O(1)$ are valid.

Finally we show the neutrino mass matrix
$M_\nu$ in each case that the neutrinos have hierarchical masses or 
degenerate masses.

In the case that the neutrinos have hierarchical masses 
($\xi_{\nu_1} \ll \xi_{\nu_2} \ll 1$), 
\begin{equation}
M_{\nu}\simeq m_{\nu_3}
\left(
 \begin{array}{ccc}
     \xi_{\nu_1} & 0 & 0 \\
     0 & 1/2 & -e^{-i\theta_1}/2  \\
     0 & -e^{-i\theta_1}/2 & e^{-2i\theta_1}/2 
\end{array}
\right) \label{36},
\end{equation}
and in the case that the neutrinos have degenerate masses 
($\xi_{\nu_1} \sim \xi_{\nu_2} \sim 1$),
\begin{equation}
M_{\nu} \simeq m_{\nu_3}
\left(
 \begin{array}{ccc}
     1 & 0 & 0 \\
     0 & e^{i\theta_1}\cos \theta_1 & i \sin \theta_1 \\
     0 & i \sin \theta_1 & e^{-i\theta_1}\cos \theta_1
\end{array}
\right) \label{37}.
\end{equation}

These simple forms (\ref{36}) and (\ref{37}) of neutrino mass matrix 
are almost same from weak scale to GUT scale, 
since the renormalization effects are small as it is possible to be 
neglected except for the special cases \cite{T}. 
The smallness of the renormalization effects is based on the 
fact that each element of the neutrino mass matrix change into 
the logarithm of the energy.

If (\ref{36}) and (\ref{37}) are originated from some symmetry of 
fundamental theory and these forms are ensured by this symmetry, 
it will become important to explore such symmetry 
as future works.

\section{Summary}

\hspace*{\parindent}
In conclusion we have proposed exact parametrization of 
the lepton mass matrices under the assumption 
that the neutrinos are Majorana particles.
We have chosen the form which has no contribution to 
$\nu_e-\nu_{\mu}$ mixing as the neutrino mass matrix $M_{\nu}$ 
and the NNI form as the charged lepton mass matrix $M_l$.
This is the exact parametrization of the lepton mass matrices, 
which reflects the small $\nu_e-\nu_{\mu}$ mixing and 
the large $\nu_{\mu}-\nu_{\tau}$ mixing.

Let us comment about the differences 
between quark and lepton case in short.
At first the number of parameters is the same as the number 
of the physical quantities in the lepton case, 
so there remains no redundant parameters 
unlike the quark case \cite{B.L.M}.
Second, we cannot transform into NNI form 
both $M_{\nu}$ and $M_l$ at the same time 
although we can do in the quark case.
Because the number of the parameters included in the mass matrices 
is smaller than the number of the physical quantities 
in the lepton case. 

Our simple parametrization may be useful to find some symmetry 
from the new physics beyond the standard model.
As future works there remains the problem to find such 
symmetry.

\vspace{20pt}
\noindent
{\Large {\bf Acknowledgement}}

\noindent
The authors would like to thank Prof. A. I. Sanda for 
useful discussions and fruitful comments.
We would like to thank Prof. Y. Sugiyama for careful reading 
of manuscript.

\end{document}